\documentstyle[12pt]{article}
\begin{document}
\title{Large Numbers, Galactic Rotation and Orbits}
\author{B.G. Sidharth$^*$\\
Centre for Applicable Mathematics \& Computer Sciences\\
B.M. Birla Science Centre, Adarsh Nagar, Hyderabad - 500 063 (India)}
\date{}
\maketitle
\footnotetext{$^*$Email:birlasc@hd1.vsnl.net.in; birlard@ap.nic.in}
\begin{abstract}
The time variation of the gravitational constant $G$ in the recently
discussed large number cosmologies accounts for the galactic rotational
velocity curves without invoking dark matter and also for effects like
the precession of the perhelion of Mercury.
\end{abstract}
In recent issues (Matthews, 1998), (Sidharth, 1999), the large number coincides brought to light
by Dirac a little over sixty years ago were revisited. Cosmological schemes
where these relations between the fundamental micro physical constants and large
scale parameters like the Hubble constant and the number of particles in
the universe are not mere magical or mysterious coincidences were considered.
Here, the universal constant for gravitation $G$ varies with time (Narlikar, 1993),
(c.f. also ref. (Sidharth, 1999)),
\begin{equation}
G = G_o(1-\frac{t}{t_o})\label{e1}
\end{equation}
where $t_o \sim 10^{17} secs$ is the present age of the universe and $t$ is
the time elapsed from the present epoch. Subscripts refer to values at the
present epoch.\\
From this it would also follow that the distance of an object moving under
the influence of a massive gravitating body would decrease with time as
(cf.ref.(Narlikar, 1993))
$$r = r_o(1-\frac{t}{t_o})$$
More accurately we have
\begin{equation}
r = r_o (1-\frac{\beta t}{t_o}) , \quad \beta \approx 1\label{e2}
\end{equation}
(There is also a variant of the above idea, called multiplicative creation
of particles, under which the distances increase with time - cf.ref.(Narlikar, 1993)
for details.)\\
We now show that given (\ref{e1}), it is possible to explain the anomalies
in galactic rotational curves on the one hand and deduce effects like the
correct perhelion precession of the planet Mercury on the other.\\
The problem of galactic rotational curves is well known (cf.ref.(Narlikar, 1993)). We
would expect, on the basis of straightforward dymamics that the rotational
velocities at the edges of galaxies would fall off according to
\begin{equation}
v^2 \approx \frac{GM}{r}\label{e1b}
\end{equation}
On the contrary the velocities tend to a value,
\begin{equation}
v \sim 300 km/sec\label{e3}
\end{equation}
This has lead to the postulation of as yet undetected dark matter, that is that the galaxies
are more massive than their visible material content indicates.\\
Our starting point is the well known equation for Keplerian orbits (Goldstein, 1966),
which on use of (\ref{e1}) becomes
\begin{equation}
\frac{1}{r} = \frac{mk_o}{l^2} (1+e cos\Theta)(1-\frac{t}{t_o}), \quad
k=GmM, l = mr^2 \dot \Theta\label{e4}
\end{equation}
$M$ and $m$ being respectively the masses of the central and orbiting objects
and $e$ is the eccentricity. From (\ref{e5}) we could deduce
\begin{equation}
r^3 \dot {\Theta}^2 = r^3_o \dot \Theta^2_o (1-\frac{t}{t_o})\label{e5}
\end{equation}
Equation (\ref{e5}), for the special case of closed orbits, $e < 1$ can be considered
to be the generalisation of Kepler's third law.\\
From (\ref{e2}) it can be easily deduced that
\begin{equation}
a \equiv (\ddot{r}_{o} - \ddot{r}) \approx \frac{\beta}{t_o} (t\ddot{r_o} + 2\dot r_o) 
\approx -2\beta \frac{r_o}{t^2_o}\label{e6},
\end{equation}
as we are considering infinitesimal intervals $t$ and nearly circular orbits.
Equation (\ref{e6}) shows that there is an anomalous inward acceleration, as if there
is an extra attractive force, or an additional central mass.\\
While we recover the usual theory, in the limit $\beta \to 0$, if we retain
$\beta$ then we will have instead of the usual equation (\ref{e1b}), in view
of (\ref{e6}) and the fact that $\beta \approx 1$
\begin{equation}
\frac{GMm}{r^2} + \frac{2mr}{t^2_o} \approx \frac{mv^2}{r}\label{e7}
\end{equation}
From (\ref{e7}) it follows that
\begin{equation}
v \approx \left(\frac{2r^2}{t^2_o} + \frac{GM}{r}\right)^{1/2}
\label{e8}
\end{equation}
From (\ref{e8}) it is easily seen that at distances within the edge of a typical
galaxy, that is $r < 10^{23}cms$ the equation (\ref{e1b}) holds but as we
reach the edge and beyond, that is for $r \geq 10^{24}cms$ we have $v \sim 10^7cms$
per second, in agreement with (\ref{e3}).\\
Thus the time variation of $G$ given in equation (\ref{e1}) explains observation
without taking recourse to dark matter.\\
We now come to the case of the precession of Mercury's perhelion. Indeed
using (\ref{e2}) in (\ref{e5}) we get
$$\dot \Theta^2 = \dot \Theta^2_o (1-\frac{t}{t_o}) (1+\frac{3 \beta t}{t_o})$$
whence,
\begin{equation}
\dot \Theta = \dot \Theta_o (1+\frac{t}{t_o})\label{e9}
\end{equation}
From (\ref{e9}) we can deduce
$$\lambda (t) \equiv \Theta - \Theta_o = \frac{\pi}{\tau_o t_o}t^2$$
where $\lambda (t)$ is the average perhelion precession at time $t$ and
$\tau_o \approx 0.25$ years the planet's period of revolution. Summing over
the years $t = 1,2,\cdots, 100,$ the total precession in a century is given by
$$\lambda = \sum^{100}_{n=1} \lambda (n) \approx 43''$$
the age of the universe $t_o$ being taken $\approx 2 \times 10^{10}$ years.
This ofcourse is the observed value.\\
Finally, it may be mentioned that several recent studies show that the
universe is ever expanding (Perlmutter, 1998), which undermines the
conventional belief that dark matter closes the universe.\\
\vspace{.25in}
\begin{flushleft}
{\bf References}\\
\smallskip
\end{flushleft}
\noindent
Goldstein H 1966  Classical Mechanics Addison-Wesley, Reading,Mass.\\
Matthews R 1998 A\& G {\bf 39} 6.19-6.20.\\
Narlikar J V 1993 Introduction to Cosmology Cambridge University Press,
Cambridge.\\
Perlmutter S et al. 1998 Nature {\bf 391} 51-54.\\
Sidharth B G 1999 A\& G {\bf 40} 2.8.
\end{document}